\documentclass[aps,prb,twocolumn,groupedaddress,showpacs,amsmath,amssymb]{revtex4}

\usepackage{graphicx}

\begin{document}

\title{Electron Localization in La$_{2-x}$Sr$_x$CuO$_4$ and the Role of Stripes}

\author{Seiki Komiya}
\email[]{komiya@criepi.denken.or.jp}
\author{Yoichi Ando}
\email[]{ando@criepi.denken.or.jp}

\affiliation{Central Research Institute of Electric Power Industry, 
Komae, Tokyo 201-8511, Japan}

\date{\today}

\begin{abstract}

The normal-state in-plane resistivity ($\rho_{ab}$) is measured in
lightly Zn-doped La$_{2-x}$Sr$_x$CuO$_4$ (LSCO) crystals with $0.06\leq
x \leq 0.17$ to systematically study the localization behavior. It is
found that the localization temperature, $T_{\rm loc}$, where $\rho_{ab}(T)$
turns from metallic ($d\rho_{ab}/dT>0$) to insulating
($d\rho_{ab}/dT<0$), shows an anomalous enhancement at $x=0.12$.
Intriguingly, the doping dependence of $T_{\rm loc}$ in Zn-doped LSCO is
found to be similar to that of Zn-free LSCO where superconductivity is
suppressed by a high magnetic field. This suggests that the mechanism of
the localization is the same in lightly Zn-doped LSCO and in Zn-free
LSCO under magnetic field, and in both cases it is probably caused by
the freezing of the electrons into an inhomogeneous state, which leads
to the spin stripes at low temperatures.

\end{abstract}


\maketitle

In underdoped La$_{2-x}$Sr$_x$CuO$_4$ (LSCO) superconductors, it is
known that the temperature dependence of the in-plane resistivity
$\rho_{ab}(T)$ becomes insulating at low temperatures when the
superconductivity is suppressed by a high magnetic
field,\cite{andoPRL95} and that the metal-to-insulator crossover in the
normal state takes place at optimum doping.\cite{boebinger} Since it is
unusual to have an insulating normal state in a superconductor, it 
would be of importance to understand the origin of this 
peculiar localization in underdoped LSCO. On the
other hand, it is also known that a metallic transport ($d\rho_{ab}/dT >
0$) is realized in LSCO with only 1\%-Sr doping at moderate
temperatures.\cite{mobility} These two peculiarities make a systematic
study of the localization behavior, the transition from high temperature
metallic $\rho_{ab}$ to low temperature insulating $\rho_{ab}$,
particularly meaningful in extending our knowledge on the electronic
state of underdoped LSCO. Since the localization behavior was studied by
the measurements in pulsed high magnetic field only for a few $x$
values,\cite{boebinger} it is the purpose of the present work to study
the $x$ dependence in more detail to understand the mechanism of the
localization. 

Another peculiar feature in underdoped LSCO is the static spin stripes
which develop at low temperatures and are observed by elastic neutron
scattering experiments.\cite{kimura1, fujitaPRB02} However, there is
still no direct evidence for the presence of charge stripes in LSCO,
although experimental results that indirectly indicate some charge
inhomogeneity are accumulating.\cite{anisotropy, AFDB, dordevic, singer}
Theoretically, various forms of charge inhomogeneity (or charge order)
have been proposed to exist in cuprates.\cite{KivelsonRMP, Kivelson,
Zhang, Zaanen, Sachdev, Auerbach, Tesanovic, Anderson, DHLee} If the
charge inhomogeneity indeed exists in underdoped LSCO, the localization
phenomena might be associated with a change in the charge dynamics due
to the formation of such a structure, as is the case with the Nd-doped
LSCO where static charge stripes are known to be stabilized and
localization takes place below the temperature of
low-temperature-orthorhombic (LTO) to the low-temperature-tetragonal
(LTT) structural phase transition.\cite{tranquada, ichikawa, noda}

In this work, we systematically study the localization behavior of
underdoped LSCO using high quality single crystals where
superconductivity is suppressed by Zn substitution. The amount of Zn
doped into the Cu site is 2\%, to avoid adversely affecting the pristine
electronic state of LSCO. In fact, it was observed by $\mu$SR
measurements that a larger amount of Zn substitution causes a
significant suppression in the magnetic correlations.\cite{watanabe} We
find that the localization temperature, $T_{\rm loc}$, where
$d\rho_{ab}/dT$ turns its sign from positive to negative with decreasing
temperature, shows an anomalous enhancement at $x=0.12$, which is close
to 1/8. Since the stripe ordering of spins is known to be pronounced at
the 1/8 carrier concentration \cite{kimura1} and this order is expected
to be triggered by some form of charge inhomogeneity,\cite{KivelsonRMP,
Kivelson, Zhang, Zaanen, Sachdev, Auerbach, Tesanovic, Anderson, DHLee}
the present result strongly suggests that the localization is caused by
the freezing of the electrons into an inhomogeneous state that triggers
the spin stripes at low temperatures.

The series of La$_{2-x}$Sr$_x$Cu$_{0.98}$Zn$_{0.02}$O$_{4}$ single
crystals ($x$ = 0.06, 0.08, 0.10, 0.12, 0.15, and 0.17 in nominal
composition) are grown by the traveling-solvent floating-zone
technique.\cite{c-axis} To compensate Cu evaporation during the growth,
we prepare raw rods with the cation ratio of La : Sr : Cu : Zn =
$(2-x):x:1:0.02$ for each $x$. Inductively-coupled plasma
atomic-emission spectroscopy (ICP-AES) is performed to determine the
cation ratio in the grown crystals. Results are summarized in Table I.
The measured $x$ values are very close to the nominal ones, and thus in
this paper we label the samples with nominal compositions for clarity. 

\begin{table}
\caption{\label{tab:table1}Results of the ICP-AES analyses for grown 
La$_{2-x}$Sr$_x$Cu$_{1-y}$Zn$_y$O$_4$ single crystals. $y=0.02$ 
for each sample in nominal composition.}
\begin{ruledtabular}
\begin{tabular}{ccccccc}
nominal $x$ &0.06&0.08&0.10&0.12&0.15&0.17\\
\hline
actual $x$ & 0.059 & 0.078 & 0.097 & 0.117 & 0.145 & 0.165 \\
actual $y$ & 0.014 & 0.015 & 0.018 & 0.016 & 0.017 & 0.017 \\
\end{tabular}
\end{ruledtabular}
\end{table}

The crystals are cut into rectangular platelets with each edge parallel
to the crystallographic axis within an error of 1$^{\circ}$ using x-ray
Laue technique. Typical sample size is 2.0 $\times$ 0.5 $\times$ 0.15
mm$^3$, where the $c$-axis is perpendicular to the platelets. The
samples are annealed at 800$^\circ$C for 40 hours in air, followed by
rapid quenching to room temperature, to remove oxygen defects in the
as-grown crystals. The temperature dependence of $\rho _{ab}$ is
measured using a standard ac four-probe method under dc magnetic field
parallel to the $c$-axis up to 18 T in the temperature range from 3.2 K
to 300 K.

\begin{figure}[b]
\includegraphics[width=7.5cm]{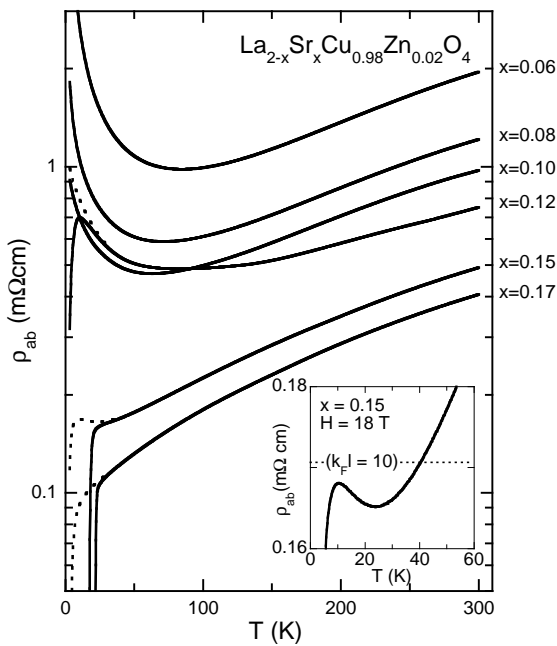}
\caption{Temperature dependences of $\rho_{ab}$ for 2\%-Zn-doped 
LSCO single crystals with $x$ = 0.06 -- 0.17. Solid lines and dotted lines 
represent the data in zero field and in 18-T field, respectively. Inset: 
Enlarged view of the data for the $x=0.15$ sample in 18-T field.}
\end{figure}

Figure 1 shows the temperature dependence of $\rho_{ab}$ in zero and
18-T fields with the vertical axis in logarithmic scale. In zero field,
except for the $x=0.12$ sample (on which we elaborate later), the
behavior of $\rho_{ab}(T)$ at moderate temperatures does not
qualitatively change with $x$, which is the same as the behavior of
Zn-free LSCO crystals.\cite{mobility} In samples with $x \leq 0.10$,
superconductivity is suppressed completely with 2\%-Zn substitution and
$\rho_{ab}(T)$ becomes insulating at low temperatures, which is also
similar to the behavior of Zn-free underdoped LSCO when
superconductivity is suppressed by pulsed high magnetic
field.\cite{andoPRL95} 

In order to additionally suppress superconductivity, 18-T magnetic field
is applied along the $c$-axis. In samples with $x \leq 0.10$, no
significant change is observed upon applying 18-T magnetic field. In the
$x=0.12$ sample, superconductivity is almost completely suppressed under
the 18-T magnetic field and $\rho_{ab}(T)$ shows an insulating behavior
down to low temperatures. In samples with $x \geq 0.15$, superconductivity
still shows up under 18 T, but we can observe a resistivity upturn below
25 K in the $x=0.15$ sample. As is shown in the inset to Fig. 1, this
resistivity upturn takes place in quite a clean system where
$\rho_{ab}\simeq170$ $\mu\Omega$cm. If we convert this resistivity into
$k_F\ell$ ($k_F$ is the Fermi wave number and $\ell$ is the mean free
path) by using the relation $k_F\ell = hc_0/\rho_{ab}e^2$ ($c_0$ is the
interlayer distance), which is derived by assuming the Luttinger theorem
and the Drude-type transport in two-dimension, we obtain $k_F\ell \simeq
10$. This value of $k_F\ell$ is unusually large for a system with localization
behavior.\cite{boebinger} On the other hand, our previous measurements
showed that $\rho_{ab}(T)$ exhibits unusually metallic behavior at
moderate temperatures in slightly hole-doped LSCO samples where $k_F\ell
\simeq 0.1$.\cite{mobility} Given that a metal (insulator) would
normally be expected for $k_F\ell \gtrsim 1$ ($k_F\ell \lesssim 1$),
these peculiar transport properties suggest that the electronic
structure in LSCO cannot be described within the conventional band
picture; we have therefore argued \cite{mobility} that a self-organized
charge inhomogeneity might be responsible for this puzzle. Note
that this unusual localization behavior is observed not only in LSCO but
also in YBa$_2$Cu$_3$O$_y$ system\cite{segawa} and in
Bi$_2$Sr$_{2-x}$La$_x$CuO$_{6+\delta}$ system.\cite{ono_PRL}

\begin{figure}
\includegraphics[width=8.6cm]{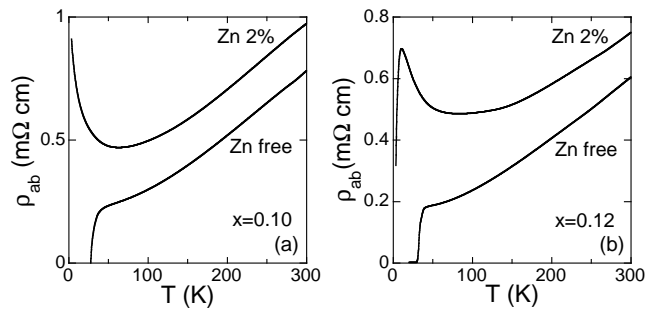}
\caption{Comparisons of $\rho_{ab}(T)$ between Zn-free and Zn-doped 
LSCO crystals at two representative dopings.}
\end{figure}

Figure 2 shows a comparison of $\rho_{ab}(T)$ in zero field for Zn-free
and Zn-doped samples at $x=0.10$ and 0.12. The data for Zn-free samples
are from our previous results.\cite{mobility, c-axis} In the normal
state, Zn-substitution usually causes temperature-independent additional
scatterings in the in-plane transport in cuprates,\cite{segawa,
fukuzumi} which is observed as a parallel shifting of $\rho_{ab}(T)$ in
the $x=0.10$ sample in Fig. 2(a). However, as is shown in Fig. 2(b), in
the $x=0.12$ sample the slope of $\rho_{ab}(T)$ changes in the Zn-doped
sample. It seems as if a resistivity of different origin grows below
$\sim$150 K in the $x=0.12$ sample. This resistivity behavior is rather
similar to that of Zn-free LSCO where superconductivity is suppressed by
high magnetic field;\cite{boebinger} namely, at $x=0.12$, an unusually
large magnetoresistance has been observed below $\sim$100 K under 60-T
magnetic field. 

To map out the localization behavior in the phase diagram, we define the
temperature where $\rho_{ab}(T)$ shows a minimum value as the
localization temperature $T_{\rm loc}$. Figure 3 shows the $x$
dependences of $T_{\rm loc}$ for 2\%-Zn-doped LSCO crystals together
with the data obtained from the previous results for Zn-free LSCO
crystals.\cite{mobility, boebinger, c-axis, komiyaLT23} The overall
tendency of $T_{\rm loc}$ in Zn-doped samples is quite similar to that
in Zn-free samples: $T_{\rm loc}$ tends to decrease with increasing $x$
and disappear at $x=0.17$, and it is clearly seen that $T_{\rm loc}$ is
anomalously enhanced at $x=0.12$ in both Zn-doped and Zn-free samples.
At this carrier concentration, the temperature where the static spin
stripes are formed is also enhanced, which was observed by elastic
neutron scattering experiments,\cite{kimura1} and thus the enhancement
of $T_{\rm loc}$ at $x=0.12$ indicates that the localization observed in
resistivity has relevance to the spin stripes. In recent $\mu$SR
measurements,\cite{watanabeMOS, panagopoulos} the temperature where the
Cu spin dynamics changes, which we call $T_0$ here, was analyzed in a
series of LSCO samples and this $T_0$ was found to show a doping
dependence similar to that of $T_{\rm loc}$. This fact further supports
the idea that the localization is related to the formation of the spin
stripes, which is reflected in the Cu spin dynamics. In passing, the
similarity of the role of Zn doping to that of high magnetic fields
demonstrated in Fig. 3 motivates one to estimate the field equivalence
of the 2\%-Zn doping: comparison of the present data to those of Refs.
\onlinecite{andoPRL95, boebinger} suggests that the 2\%-Zn doping
roughly corresponds to 40 T.\cite{note}

\begin{figure}
\includegraphics[width=7.5cm]{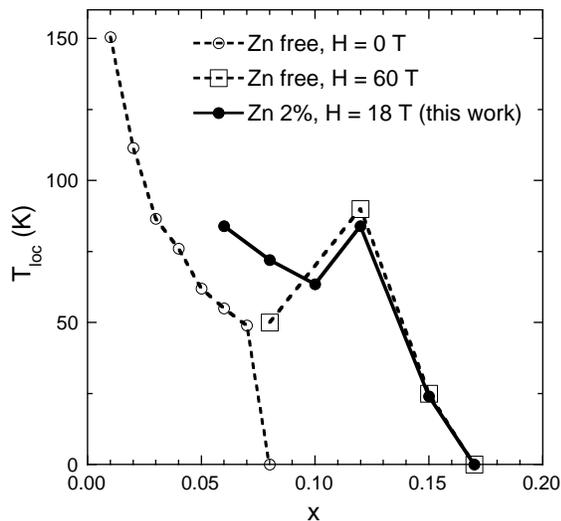}
\caption{$x$ dependences of $T_{\rm loc}$ for Zn-free and 2\%-Zn-doped
LSCO single crystals. Open circles and squares denote $T_{\rm loc}$ of
Zn-free LSCO. \cite{boebinger, mobility, c-axis, komiyaLT23} Solid
circles represent $T_{\rm loc}$ of Zn-doped LSCO (this work); note that
the 18-T field affects $T_{\rm loc}$ only in the $x$ = 0.15 samples.}
\end{figure}

There are other results which we consider to support the relation
between the localization and the spin texturing. Lake {\it et al.}
\cite{lakeNature} observed an enhancement of the incommensurate magnetic
peaks in underdoped LSCO by magnetic field in their elastic neutron
scattering experiments. Correspondingly, Sun {\it et al.} \cite{sunPRL}
and Hawthorn {\it et al.} \cite{hawthornPRL} observed decreasing
quasiparticle thermal conductivity with increasing magnetic field in
underdoped LSCO crystals at very low temperatures. Taken together, these
results indicate that a magnetic-field-induced localization of
quasiparticles \cite{sunPRL} is associated with a magnetic-field-induced
spin order, and this quasiparticle localization in the superconducting
state is probably responsible for the insulating behavior under high
magnetic field.

It is useful to note that the antiferromagnetism is competing with
superconductivity in cuprates,\cite{Zhang,Sachdev} and both the magnetic
field \cite{lakeNature} and the Zn doping \cite{Alloul} enhance the
antiferromagnetism; in this sense, the similarity between the roles of
the two is rather natural. Also, it is widely conjectured
\cite{KivelsonRMP, Kivelson, Zhang, Zaanen, Sachdev, Auerbach,
Tesanovic, Anderson, DHLee} that the magnetic state that is competing
with superconductivity involves some texturing of spins
\cite{lakeNature} and charges.\cite{Hoffman, Yazdani} Our data point to
an intimate link between the spin stripes and the charge localization,
which can be naturally understood if some charge inhomogeneity triggers
both the spin stripes and the localization. Intriguingly, a
charge-inhomogeneous state has certain similarities to a granular system,
for which the $\log(1/T)$ resistivity has been experimentally observed
\cite{Wolf} and theoretically proposed \cite{Vinokur}. Thus, we
speculate that the peculiar localization in underdoped cuprates may
originate from an effective granularity brought about by some form of
charge inhomogeneity when the superconductivity is suppressed.

In passing, one can see in Fig. 3 that $T_{\rm loc}$ of 2\%-Zn-doped
LSCO with $x>0.10$ is almost identical to that of Zn-free LSCO under
high magnetic field, while for $x<0.10$, $T_{\rm loc}$ of Zn-doped
samples is about 20 K higher than that of Zn-free ones. This difference
would be related to the Fermi-surface topology elucidated by the
angle-resolved photoemission spectroscopy (ARPES)
measurements:\cite{teppei} In slightly Sr-doped LSCO samples, only the
Fermi arc is observed at the Fermi energy $E_F$ near the ($\pi/2$,
$\pi/2$) position in the Brillouin zone, and electronic states near
($\pi$, 0) are located deeply below $E_F$,\cite{ino} which suggests that
the electronic transport is governed by the quasiparticles on the Fermi
arcs in slightly hole-doped LSCO crystals.\cite{Hall_2band} As $x$
increases, the ($\pi$, 0) band moves up and touches $E_F$ at
$x\sim 0.10$,\cite{ino} above which the electrons near ($\pi$, 0) will
participate in the transport. Therefore, if the Zn-induced
scatterings of the quasiparticles on the Fermi arcs are stronger than
those on the ($\pi$, 0) band, the localization effect in $x<0.10$
samples, where the quasiparticles are confined on the Fermi arcs, would
be more pronounced. Further experiments would be desirable to clarify
the details of the possible scattering-rate anisotropy.

\begin{figure}
\includegraphics[width=8cm]{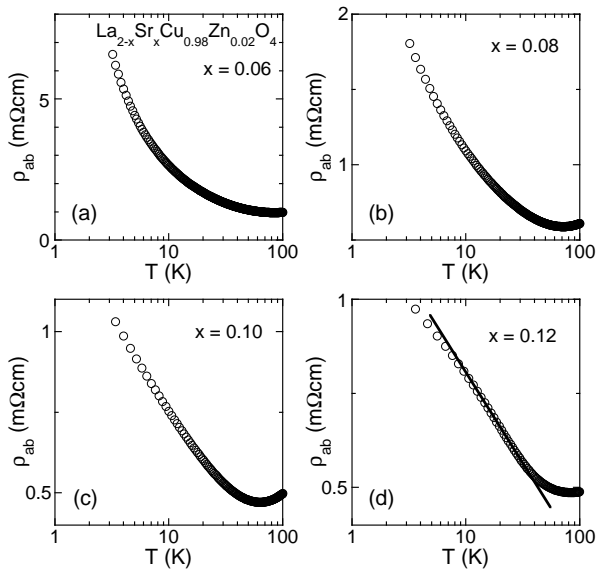}
\caption{log$(T)$ plots of $\rho_{ab}(T)$ for Zn-doped LSCO with $x \leq 0.12$.}
\end{figure}

Finally, let us discuss the temperature dependence of the insulating
behavior. Figure 4 shows log($1/T$) plots of $\rho_{ab}(T)$ of Zn-doped
LSCO with $x\leq0.12$. One can see the log($1/T$) insulating behavior in
the $x=0.12$ sample (where the 18-T field seems insufficient to suppress
superconductivity completely below 10 K), while for $x\leq0.10$ the
resistivity diverges faster than log($1/T$). On the other hand, the
transport study under 60-T field showed that in Zn-free LSCO at
$x=0.08$, the resistivity exhibits clear log($1/T$)
divergence.\cite{andoPRL95} Therefore, for the same Sr content, the
insulating behavior is slightly different in the two cases. It is useful
to note that in strongly insulating samples where $\rho_{ab}$ exceeds a
rough criterion of 2--3 m$\Omega$cm (which corresponds to $k_{F}\ell
\sim$ 1), $\rho_{ab}(T)$ becomes roughly consistent with the variable
range hopping (VRH) behavior; actually, the data for $x$ = 0.06 [Fig.
4(a)] is consistent with the VRH behavior. On the other hand, the
$\log(1/T)$ behavior is observed whenever the resistivity is less than
the criterion in Zn-free underdoped samples. The 2\%-Zn-doped samples at
$x$ = 0.08 and 0.10 [Figs. 4(b) and 4(c)] are peculiar in that their
$\rho_{ab}(T)$ behavior is neither $\log(1/T)$ nor VRH, even though
their resistivity is smaller than the criterion. If the $\log(1/T)$
behavior is associated with an effective granularity as we conjectured
above, the slight anomaly in the Zn-doped samples may indicate that the
Zn impurities not only stabilize the spin/charge texturing but also
disturb the local electronic states.

In summary, the insulating behavior of underdoped LSCO is studied in a
series of single crystals where superconductivity is suppressed by
2\%-Zn doping. It is found that the localization temperature, $T_{\rm
loc}$, is anomalously enhanced at $x=0.12$. Intriguingly, the $x$
dependence of $T_{\rm loc}$ is found to be similar to that of Zn-free
LSCO under high magnetic fields. This suggests that the localization
mechanism is essentially the same in Zn-doped LSCO and in Zn-free LSCO
under magnetic field. Furthermore, the behavior of $T_{\rm loc}$ is
rather similar to that of the spin freezing temperature determined by
$\mu$SR measurements. Based on these results, we discuss possible
relevance of the spin/charge texturing to the localization phenomena. 

We thank M. Strongin for helpful discussions.


\begin{thebibliography}{99}
\bibitem{note}
For example, the Zn-free $x$ = 0.15 sample
at 4.2 K is in the middle of the transition in 40 T;\cite{boebinger}
correspondingly, the 2\%-Zn-doped $x$ = 0.15 sample is in the
middle of the transition at 4.2 K.
\end{thebibliography}
\end{document}